# In which fields do higher impact journals publish higher quality articles?

Mike Thelwall
Statistical Cybermetrics and Research Evaluation Group, University of Wolverhampton, UK.
https://orcid.org/0000-0001-6065-205X m.thelwall@wlv.ac.uk

Kayvan Kousha
Statistical Cybermetrics and Research Evaluation Group, University of Wolverhampton, UK.
https://orcid.org/0000-0003-4827-971X k.kousha@wlv.ac.uk

Meiko Makita
Statistical Cybermetrics and Research Evaluation Group, University of Wolverhampton, UK.
https://orcid.org/0000-0002-2284-0161 meikomakita@wlv.ac.uk

Mahshid Abdoli
Statistical Cybermetrics and Research Evaluation Group, University of Wolverhampton, UK.
https://orcid.org/0000-0001-9251-5391 m.abdoli@wlv.ac.uk

Emma Stuart
Statistical Cybermetrics and Research Evaluation Group, University of Wolverhampton, UK.
https://orcid.org/0000-0003-4807-7659 emma.stuart@wlv.ac.uk

Paul Wilson
Statistical Cybermetrics and Research Evaluation Group, University of Wolverhampton, UK.
https://orcid.org/0000-0002-1265-543X pauljwilson@wlv.ac.uk

Jonathan Levitt
Statistical Cybermetrics and Research Evaluation Group, University of Wolverhampton, UK.
https://orcid.org/0000-0002-4386-3813 j.m.levitt@wlv.ac.uk

**Acknowledgement**
This study was funded by Research England, Scottish Funding Council, Higher Education Funding Council for Wales, and Department for the Economy, Northern Ireland as part of the Future Research Assessment Programme (https://www.jisc.ac.uk/future-research-assessment-programme). The content is solely the responsibility of the authors and does not necessarily represent the official views of the funders.

**AUTHOR CONTRIBUTIONS**
Mike Thelwall: Methodology, Writing–original draft, Writing–review & editing.
Kayvan Kousha: Methodology, Writing–review & editing.
Emma Stuart, Meiko Makita, Mahshid Abdoli, Paul Wilson, Jonathan Levitt: Writing–review & editing.

**COMPETING INTERESTS**
Thelwall and Kousha are members of the distinguished reviewers board of Scientometrics.

# In which fields do higher impact journals publish higher quality articles?

The Journal Impact Factor and other indicators that assess the average citation rate of articles in a journal are consulted by many academics and research evaluators, despite initiatives against overreliance on them. Despite this, there is limited evidence about the extent to which journal impact indicators in any field relates to human judgements about the journals or their articles. In response, we compared average citation rates of journals against expert judgements of their articles in all fields of science. We used preliminary quality scores for 96,031 articles published 2014-18 from the UK Research Excellence Framework (REF) 2021. We show that whilst there is a positive correlation between expert judgements of article quality and average journal impact in all fields of science, it is very weak in many fields and is never strong. The strength of the correlation varies from 0.11 to 0.43 for the 27 broad fields of Scopus. The highest correlation for the 94 Scopus narrow fields with at least 750 articles was only 0.54, for Infectious Diseases, and there was only one negative correlation, for the mixed category Computer Science (all). The results suggest that the average citation impact of a Scopus-indexed journal is never completely irrelevant to the quality of an article, even though it is never a strong indicator of article quality.

## 1 Introduction

Formulas to calculate the average citation rate of articles in a journal, such as the Journal Impact Factor (JIF) were originally designed to help academics to find important journals in their field (Garfield, 1972). This is reasonable on the basis that journals attracting more citations are more likely to contain articles that could be cited and may even tend to publish more useful or better articles, other factors being equal. JIFs and similar journal rankings have since been used to evaluate researchers (McKiernan et al., 2019) and have become targets for academics seeking to publish in the most prestigious outlets (e.g., Salandra et al., 2021; Śpiewanowski & Talavera, 2021; Walker et al., 2019) or get recognition for their work (Brooks et al., 2021). This is logical because publishing in higher ranked journals associates with career success in some fields (e.g., finance: Bajo et al., 2020). In fields where JIFs are valued, they may generate a positive feedback loop, where authors attempt to get their best work into journals with the highest JIFs. This may even change the field by encouraging authors to standardise on research conforming to the expectations of reviewers for the high impact journals (e.g., Kitayama, 2017). In contrast, citations and all citation-based indicators may be meaningless and unvalued in some areas of academia, such as the arts and humanities (Thelwall & Delgado, 2015). Thus, it seems likely that journal-level impact evidence may have value in some academic fields, but not others.

    A focus on journal impact can have clearly negative effects by pushing academics away from their preferred publishing styles, research topics (Brooks et al., 2021) and locally-relevant research (Lee & Simon, 2018). It can also undervalue less cited specialties through their associated journals (Stockhammer et al., 2021). The widespread misuse of JIFs and similar journal-level indicators led to initiatives to restrict their use in evaluation, such as the San Francisco Declaration of Research Assessment (DORA, 2020), which is now accepted in the UK (UKRI, 2020). DORA emphasises that the value of an article should not be reduced to

the value of the publication venue. Nevertheless, the continued importance of JIFs for academics is suggested by their prominent appearance on many journal websites, except where there are agreements to avoid them (e.g., Casadevall et al., 2016).

## 1.1  *Journal impact and journal quality rankings*

From an evaluation perspective, journal impact indicators have the advantage that in some fields they seem to broadly reflect quality hierarchies, for example with high values for prestigious journals like *Cell*, *Lancet* and *NEJM*. In addition, they are relatively objective and draw on many individual academic decisions by editors, reviewers, and citing authors (Waltman & Traag, 2020). For example, the Association of Business Schools (ABS) journal ranking list, which is composed by subject experts, uses JIFs to support the judgements needed (Kelly et al., 2013; see also: charteredabs.org/academic-journal-guide-2021/), reflecting a belief that they have value but are imperfect within business. National journal ranking lists are sometimes also informed by JIFs (e.g., Pölönen et al., 2021). Journal rankings constructed by experts have their own flaws because academics tend to give higher ratings to journals in their own field (Serenko & Bontis, 2018) and the results vary by country (Taylor & Willett, 2017).

Previous studies have mainly assessed the value of journal impact factors either from a theoretical perspective or with data from a single field. A 2016 systematic review found 18 articles that had correlated JIFs with expert judgements of journal value in science, technology, and social science fields (Mahmood, 2017; Table 2).  The sample sizes were small, ranging from 8 to 127 journals. The Spearman or Pearson correlation coefficients ranged from -0.112 (regional science, n=70) to 0.836 (risk management and insurance: n=13). The low sample sizes make it difficult to draw strong conclusions, however. For example, 95% confidence intervals for the two negative correlations (e.g., Maier, 2006) both include zero so there is insufficient evidence to conclude that there is an underlying negative relationship between journal impact and prestige in any field. Nevertheless, the results suggest moderate or strong correlations in business-related fields, including management sciences (r=0.77, n=39), decision and management sciences (r=0.47, n=47), risk management and insurance (r=0.836, n=13), finance (r=0.43, n=29), and environmental and resource economics (rho=0.59, n=11). They also suggest moderate or strong correlations in health-related fields, including internal medicine (r=0.82, n=9), clinical neurology (r=0.67, n=41), diabetes (r=0.48, n=20), ophthalmology (rho=0.65, n=28), and biomechanics (r=0.35, n=46). The social sciences investigated had weak or moderate correlations: social work (rho=0.45, n=32), library and information science (r=0.528, r=0.267, n=71), planning (r=0.02, n=35), and safety (rho=0.33, n=19). Other broad areas of science were represented only by individual fields: statistics (r=0.56, n=54), regional science (r=0.112, n=70), artificial intelligence (rho=0.51, n=127), and design (r=0.10, n=8).

Subsequent studies have found strong correlations between JIFs and expert judgements of journals in industrial and organizational psychology (rho=0.71, n=34) (from Table 5 of: Highhouse et al., 2020). One analysis of the reputations of journals in 20 fields according to faculty in one college found substantial differences in the apparent importance of JIFs for journals, being most important in management and least in radiological and health professions (Walters & Markgren, 2019). This used an unusual research design, however, with a binary expert judgement rather than a ranking, so its results are not comparable to other studies.

A large-scale study compared Excellence in Research for Australia (ERA) 2010 expert-based rankings (four tiers) for 20,712 journals with Elsevier's Source Normalized Impact per Paper (SNIP) and the Clarivate JIF, both from 2010, organised into the 27 Scopus broad fields (Fig 1 of: Haddawy et al., 2016, see also: Haddow & Genoni, 2010). Although confidence intervals were not provided, the sample sizes were relatively large. There were positive Spearman correlations between the expert rankings and SNIPs for all 27 fields, varying from 0.28 (Arts & Humanities) to 0.74 (Dentistry and Veterinary Science). The correlations were at least 0.5 for all fields except Nursing, General, Social Sciences and Arts & Humanities. This is not a perfect test because ERA rankings seem to be influenced by JIFs (Haslam & Koval, 2010) and may reflect the prestige of the journal rather than the average value of the articles in it. ERA no longer ranks journals, but other countries do (e.g., Finland: Saarela & Kärkkäinen, 2020). All of the above studies have compared journal reputation with JIFs rather than article quality with JIFs, and journal reputations were probably influenced by JIFs in most of the fields examined.

The relationship between JIFs and article quality has also been investigated, giving direct evidence of this core issue. There is some REF evidence about the value of journal impact factors as evidence of article quality in different UoAs, although not in a peer reviewed report. An analysis of 19,130 individual output scores for the last REF (Wilsdon et al., 2015ab) found moderate statistical associations between article quality and journal citation rates (Elsevier's Source Normalised Impact per Paper [SNIP], a field normalised variant of the JIF) in medicine and physical science UoAs, with the highest being for Economics and Econometrics (Spearman correlation: 0.67). In engineering, the social sciences, and the arts and humanities, there were weak or negligible associations (Table A18 of: Wilsdon, et al., 2015b). Since confidence intervals for negative values for these data always contain zero (Figure 1), there is insufficient evidence to claim an underlying negative association in any field. This analysis used selected UK-authored journal articles only.

Finally, two large scale studies have compared expert rankings with expert quality scores for journal articles from Italy. They used a regression approach for Italian articles about architecture, arts and humanities, history, geography, philosophy, law, sociology, anthropology, education, library sciences and political sciences with article quality scores from the Italian research assessment exercise (VQR). The studies found positive associations between expert scores of articles and expert rankings of these journals in all these fields (Ferrara & Bonaccorsi, 2016; Bonaccorsi et al., 2015). Thus, it seems that higher rated journals tended to publish higher rated articles, when both ratings are made independently by human experts, although the association was not strong. A different analysis of the VQR data found variations between social sciences and humanities fields in the extent to which journals ranked highly by the Italian National Agency for the Evaluation of the University and Research Systems (ANVUR) tended to contain higher quality articles. The weakest evidence occurred for antiquities, philology, literary studies, and art history, and the strongest for economics and statistics (Cicero & Malgarini, 2020).

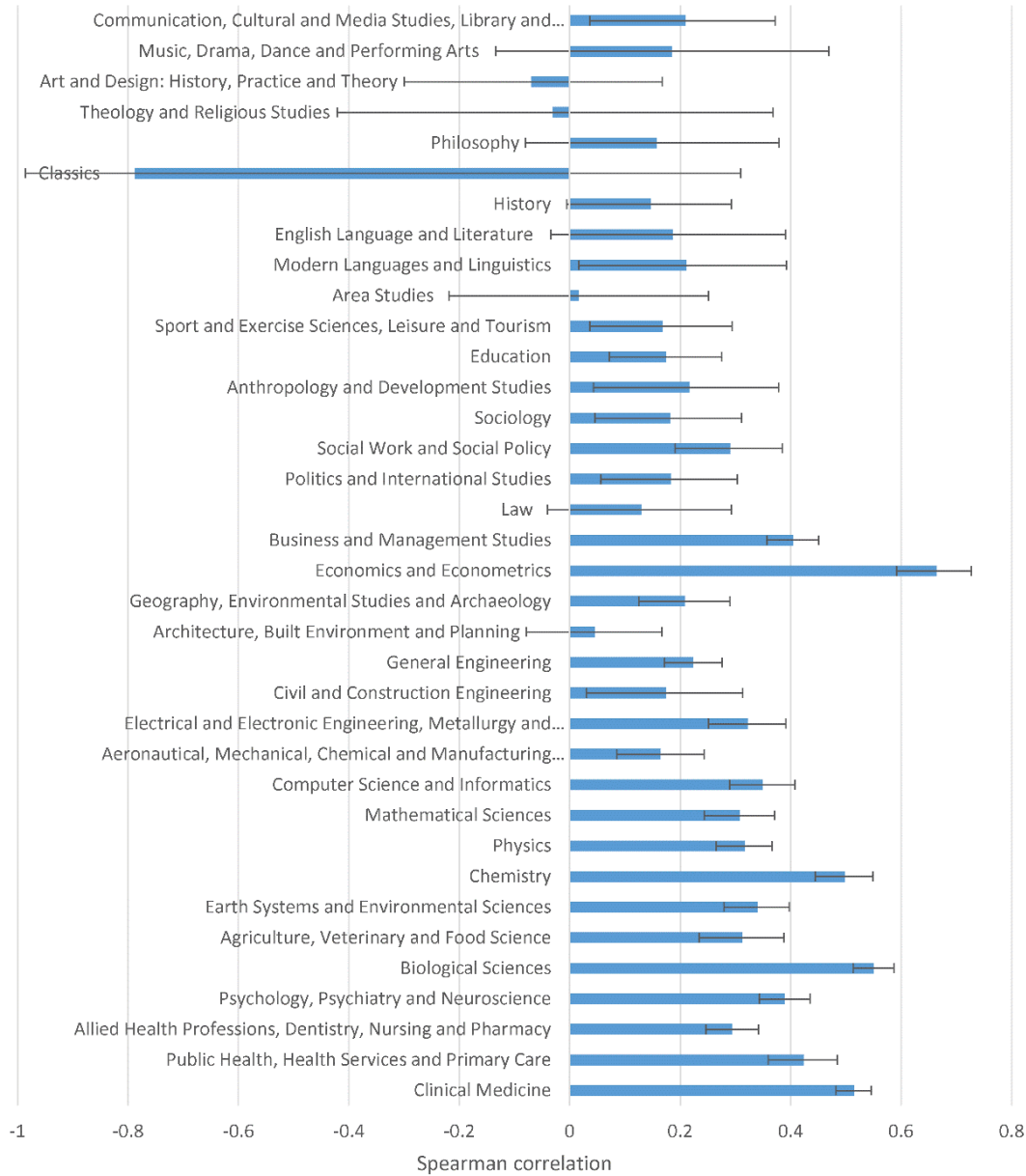

Figure 1. Spearman correlations between SNIP values and reviewer REF2014 scores for articles published in 2008 by Unit of Assessment (Table A18 of: Wilsdon, et al., 2015b). Error bars illustrate 95% confidence intervals (from the Fisher z-transformation: Fisher, 1921): wide confidence intervals are caused by small sample sizes.

## 1.2 Research questions

With the exception of the one non peer-reviewed report (Wilsdon, et al., 2015b), there are no science-wide analyses of the relationship between journal article quality and journal citation rates. This is a critical gap for those using JIFs or similar in a formal or informal research evaluation context. The current article fills this gap by replicating the earlier study (i.e., Table A18 of: Wilsdon et al., 2015b) with a more recent set of journal articles, combining years to give more precise information (narrower confidence intervals), and using a journal citation rate calculation that reduces the influence of individual highly cited articles and ties the citation rates to the publication years of the articles. Two research questions are addressed; the second concerns the field categorisation scheme because many research evaluation exercises use their own categories so it is important to assess how this might affect the results.
- RQ1: To what extent are higher quality articles published in higher impact journals?
- RQ2: How does the answer to RQ1 depend on the field categorisation scheme used?

# 2 Methods

As stated above, we replicated a previous study (Wilsdon et al., 2015b), with more recent data, years merged for increased statistical power, and a more robust method of calculating field normalised journal impact.

## 2.1 *Data: REF scores and average journal impact*

We obtained provisional REF scores for REF2021 from UK Research and Innovation (UKRI) in March 2022 for 148,977 journal articles from all 34 Units of Assessment (UoAs) in the REF, organised into four main panels (A, B, C, D). For confidentiality reasons, scores for the University of Wolverhampton were removed.

All REF2021 scores were decided by expert peer review, over 1000 field specialists mainly from the UK organised into sub-panels (UoAs) (REF, 2017). Each individual article score was primarily decided by at least two independent reviewers, then ratified by the UoA sub-panel. There were also procedures to increase scoring consistency within UoAs. Scores are on the five-point scale 0, 1*, 2*, 3*, or 4*, with 0 ("falls below the standard of nationally recognised work" or "does not meet the published definition of research for the purposes of this assessment.") being the lowest, and 4* ("world-leading in terms of originality, significance and rigour") being the highest (REF, 2019). Differently from REF2014, all research active staff had to submit publications (so there may be more lower quality outputs than before) and the average number of publications per submitted full-time academic is 2.5, rather than 4. There is also more flexibility in the number of outputs submitted in REF2021. For this article, we analysed only journal articles published in Scopus-indexed journals. In arts and humanities UoAs, these were a small minority of the works submitted. Review articles are ineligible for the REF and so are not included.

We removed all 318 articles with a score of 0 (before any correlation calculations had been performed) because these seemed to sometimes indicate that the author had been disqualified. We then matched each journal article with a DOI to an article in Scopus 2014-20 (n=133,218) and matched by title/year/authors and manually checked (n=997) if no DOI match could be found. Were only retained articles with a Scopus publication date 2014-18 for analysis to give at least two years of citation data to estimate the contemporary impact of the publishing journal (n=96,031). We derived the Scopus articles from a collection that we had downloaded in January 2021 using the Scopus API.

We calculated the average citation impact of the journal publishing each article as follows. First, we (natural) log-transformed all citation counts using log(1+x) to reduce the influence of individual highly cited articles. Without this step, individual journal averages could be dominated by individual articles, giving an unrepresentative result (Thelwall & Fairclough, 2015). Next, we calculated the average log-transformed citation count for all journal articles separately for each Scopus narrow field and year. We then divided each article log-normalised citation count by the average for the field and year containing it, giving a Normalised Log-transformed Citation Score (NLCS) (Thelwall, 2017). The NLCS for each article is therefore the ratio of its log-transformed citation count to the average log-transformed citation count for all articles in its field and year. Articles classified into multiple fields were instead divided by the average of the average log-transformed citation count for these multiple fields. The journal Mean Normalised Log-transformed Citation Score (MNLCS) is then the average of the NLCS of all articles (REF and non-REF) in the journal. This figure is independent of field and year, by design. In particular, a journal MNLCS of 1 indicates that the journal's articles tend to get an average number of citations for whichever field(s) and year in which they were published. Higher values (>1) indicate an above world average citation rate and lower values (<1) indicate a below world average citation rate.

## 2.2 Analysis

We assessed the relationship between average article quality and average citation rate using Spearman correlations. Although we had normalised the citation data to reduce skewing and the article quality data has a limited range, using Spearman instead of Pearson correlations is a conservative strategy because REF scores are ranks on a short scale (in contrast to the 27-level naturally linear VQR scores, for example). We did not use regression (in contrast to: Ferrara & Bonaccorsi, 2016; Bonaccorsi et al., 2015) because the goal is to identify overall associations between article quality and journal impact, rather than factors that influence them or that are associated with them.

We compared results from three categorisation schemes to address RQ2. The 34 REF2021 units of assessment are organised primarily to match UK academic departments and vary greatly in size. The Scopus 27 broad field scheme is a widely known standard set of categories that is used by Elsevier and others for research evaluation processes. The Scopus 334 narrow field (exact numbers varying slightly by year) scheme allows a much finer-grained comparison. For the narrow field scheme, we used a minimum of 750 articles per category as a simple threshold to reduce the results to a manageable set. Both Scopus schemes are primarily journal-based (like the Web of Science) in the sense that most articles are assigned to categories based on the publishing journal. Unlike the REF UoAs, Scopus uses a multiple category approach in which each journal (and hence each article) is usually assigned to multiple relevant broad and narrow categories.

## 3 Results

The Spearman correlations between average journal impact (MNLCS) and REF score (1* to 4*) for REF articles matching Scopus journal articles 2014-18 are positive for all UoAs, although the 95% confidence intervals contain 0 in four cases (Figure 2). The corelations tend to be very low for Main Panel D (mainly arts and humanities), with the unexpected exception of History. The correlations tend to be highest for Main Panel A (mainly health and life sciences). There are large variations within Main Panels B, C and D.

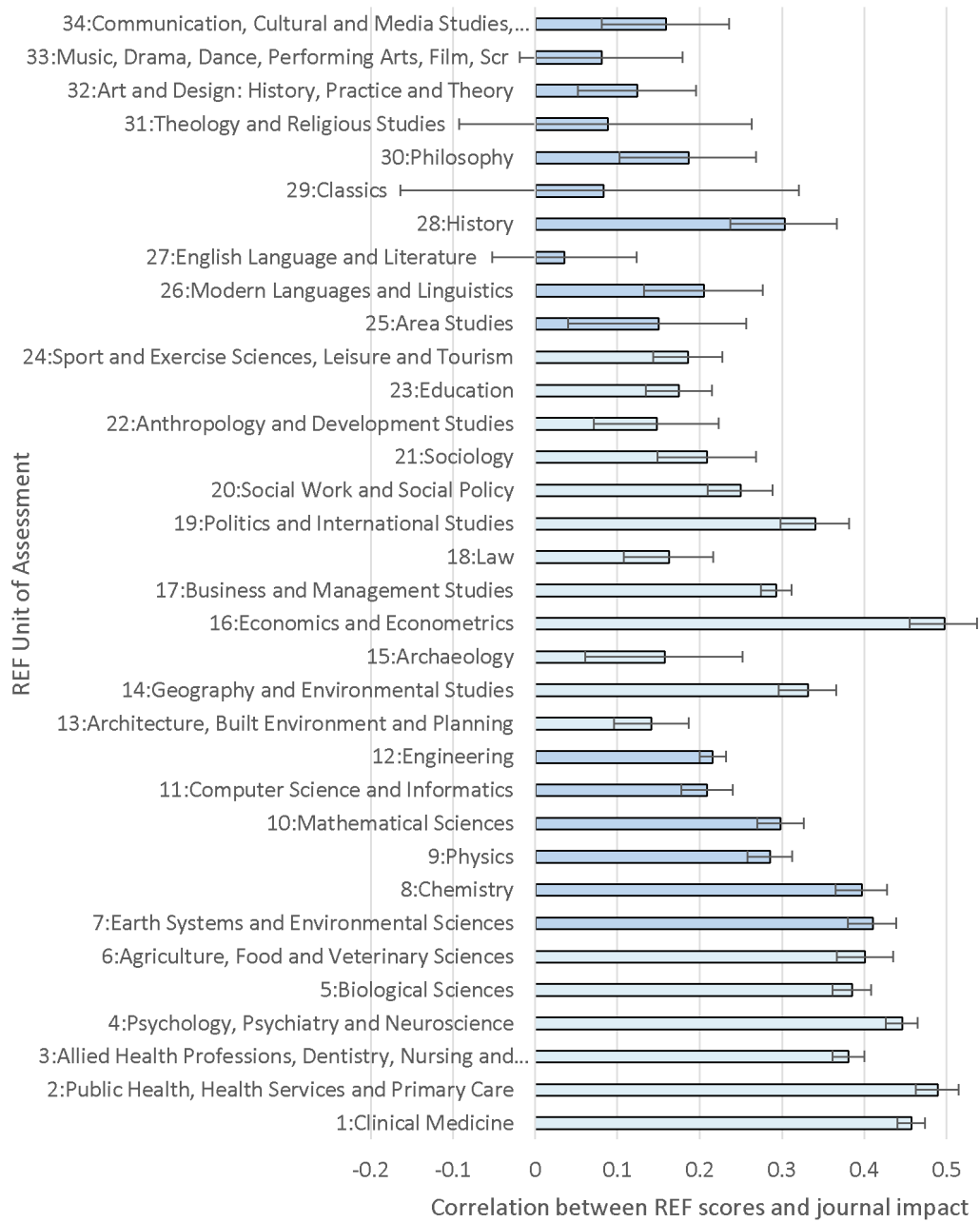

Figure 2. Article-level Spearman correlations by **UoA** between average journal impact (MNLCS) and UK REF provisional score for UK REF2021 articles matched with a Scopus journal article published 2014-18 (n=96,031). Error bars illustrate 95% confidence intervals. Slight colour changes indicate main panels A, B, C, D.

For the 27 Scopus broad fields, the correlations between REF scores and journal impact are above 0.1 in all fields and only the Veterinary confidence interval contains 0 (Figure 3). There are large variations within each of the four Scopus top-level categories (Health Sciences, Life Sciences, Physical Sciences, Social Sciences). Combined with the UoA results above, this confirms that there is not a simple disciplinary rule about the types of scholarship in which journal impact associates most strongly with article quality.

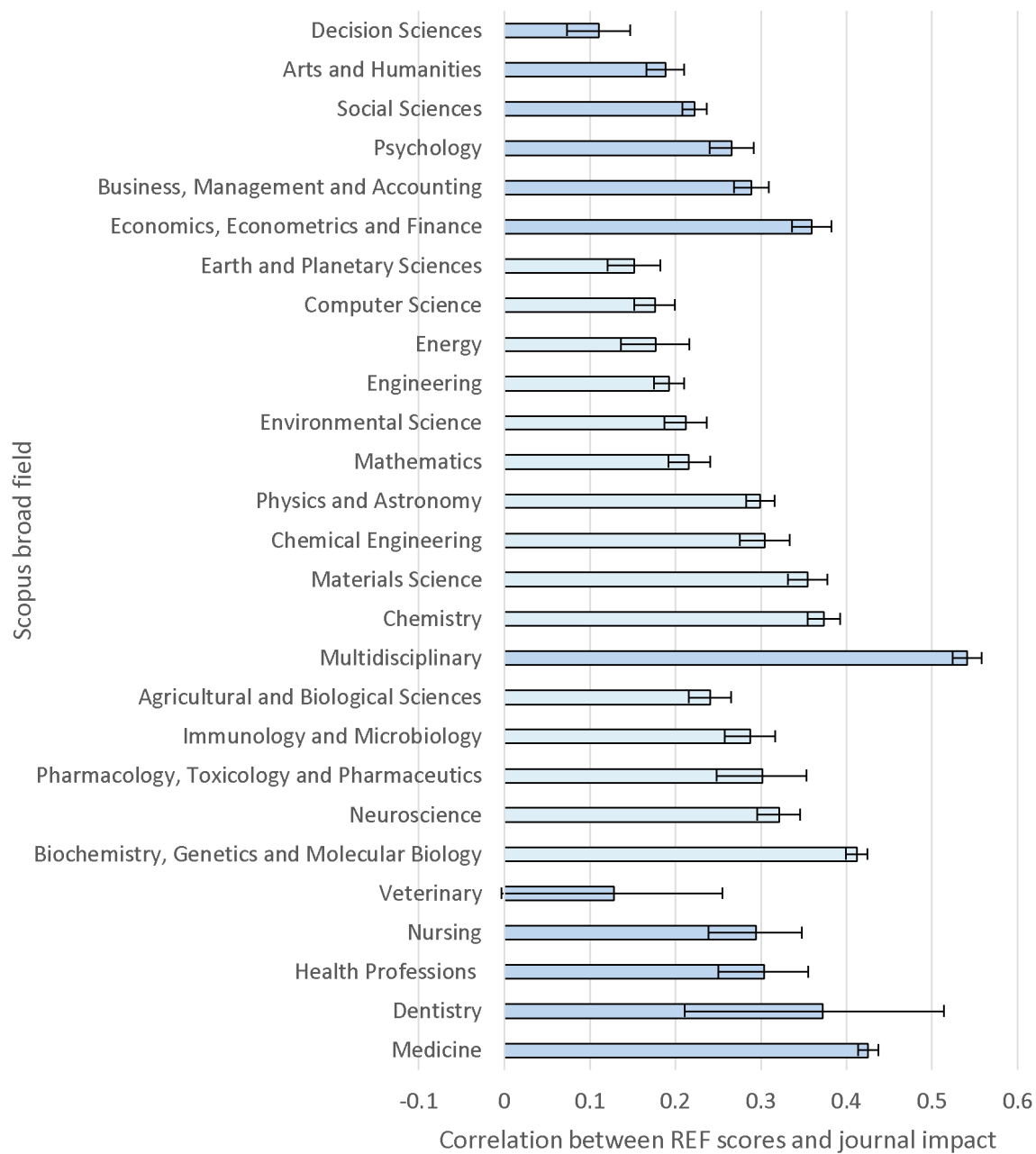

Figure 3. Spearman correlations by Scopus **broad field** between average journal impact (MNLCS) for UK REF provisional score for UK REF2021 articles matched with a Scopus journal article published 2014-18 (n=169,555, double counting articles in multiple broad fields). Broad fields are ordered by correlation within the four Scopus Top-level areas, plus Multidisciplinary (indicated by slight colour changes). Error bars illustrate 95% confidence intervals.

Almost all Scopus narrow fields with at least 750 REF articles 2014-18 have a positive correlation between REF scores and journal impact (MNLCS) (Figures 4-7). The only exception is Computer Science (all) (Figure 6). This is one of the two unusual types of narrow field in Scopus. The "all" and "miscellaneous" narrow fields that occur within its broad fields are not narrow academic fields but are instead categories that capture articles that do not fit neatly within narrow fields. Thus, the correlations are positive for all genuine narrow fields in Scopus.

The almost universally positive narrow field correlations add weight to the evidence that journal impact associates with article quality at least a small amount in all areas of scholarship. The arts and humanities can still be an exception, however, since the numbers were small in arts and humanities fields because most REF submissions in these areas were not journal articles.

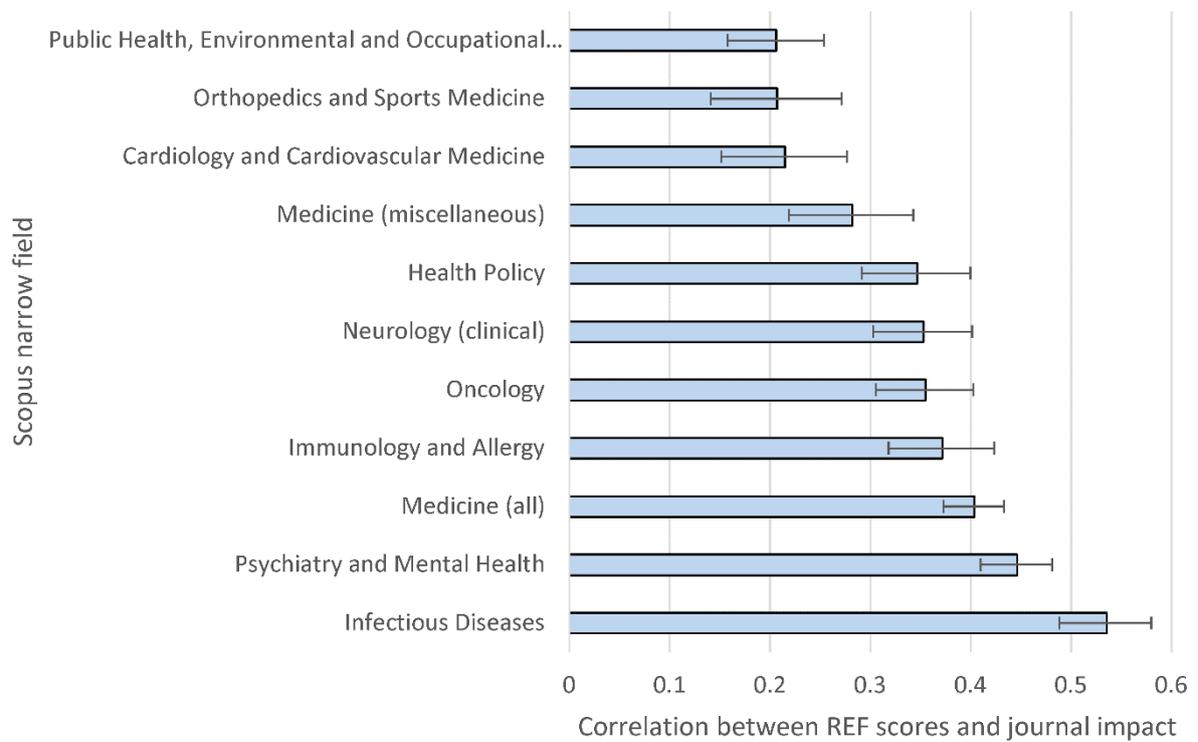

Figure 4. Spearman correlations by **narrow field** between average journal impact (MNLCS) for UK REF provisional score for UK REF2021 articles matched with a Scopus journal article published 2014-18 within a **Health Sciences** broad field. Error bars illustrate 95% confidence intervals. Qualification: At least 750 articles with REF scores.

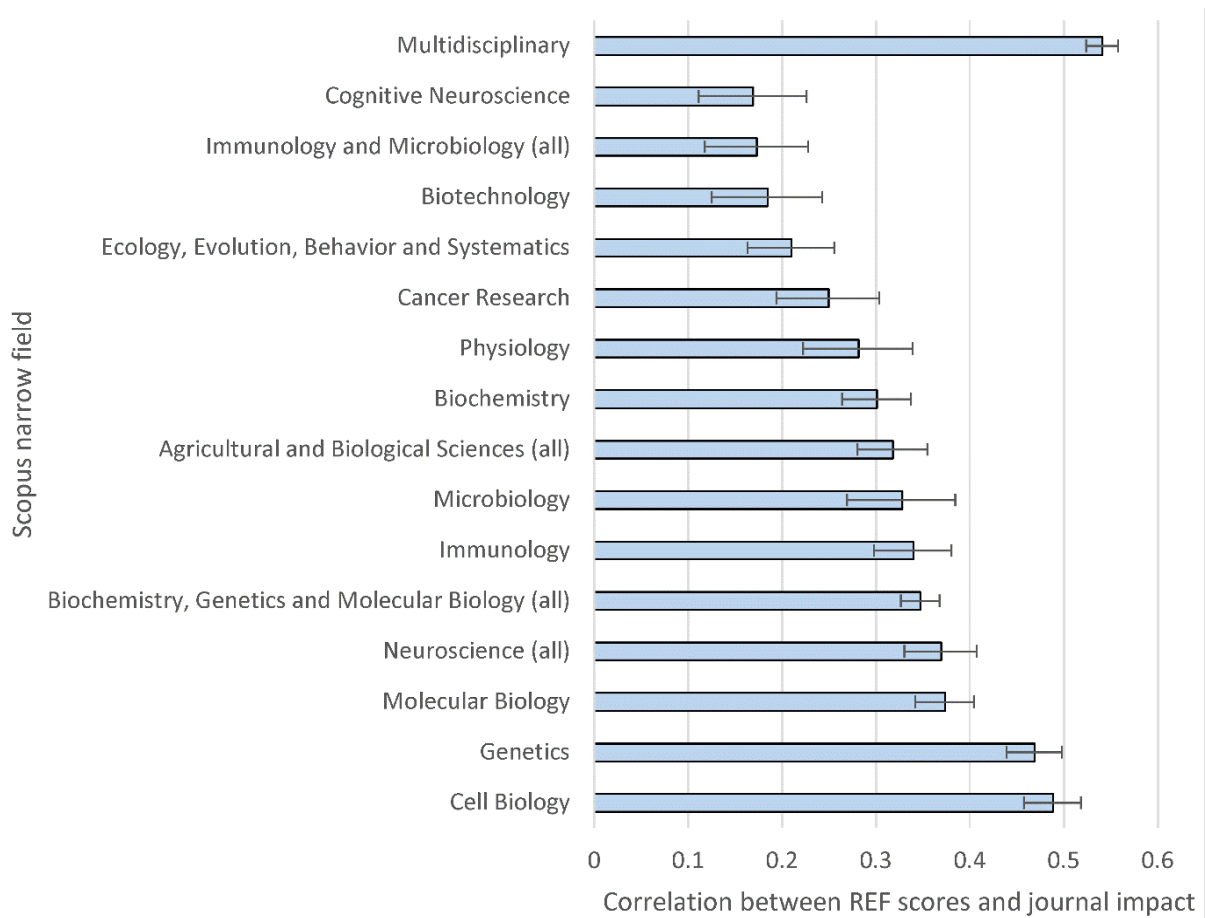

Figure 5. Spearman correlations by **narrow field** between average journal impact (MNLCS) for UK REF provisional score for UK REF2021 articles matched with a Scopus journal article published 2014-18 within a **Life Sciences** broad field. Error bars illustrate 95% confidence intervals. Qualification: At least 750 articles with REF scores.

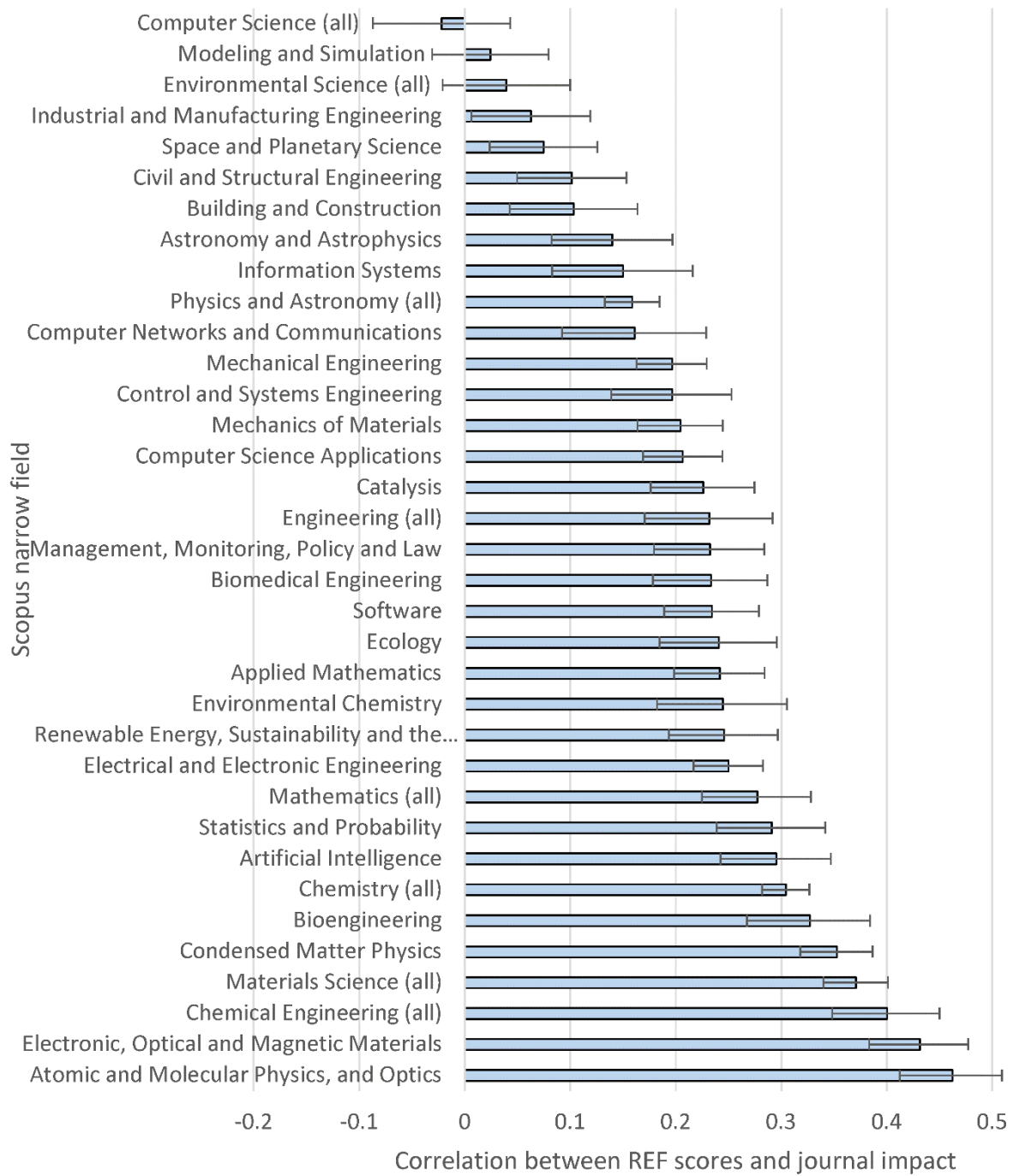

Figure 6. Spearman correlations by **narrow field** between average journal impact (MNLCS) for UK REF provisional score for UK REF2021 articles matched with a Scopus journal article published 2014-18 within a **Physical Sciences** broad field. Error bars illustrate 95% confidence intervals. Qualification: At least 750 articles with REF scores.

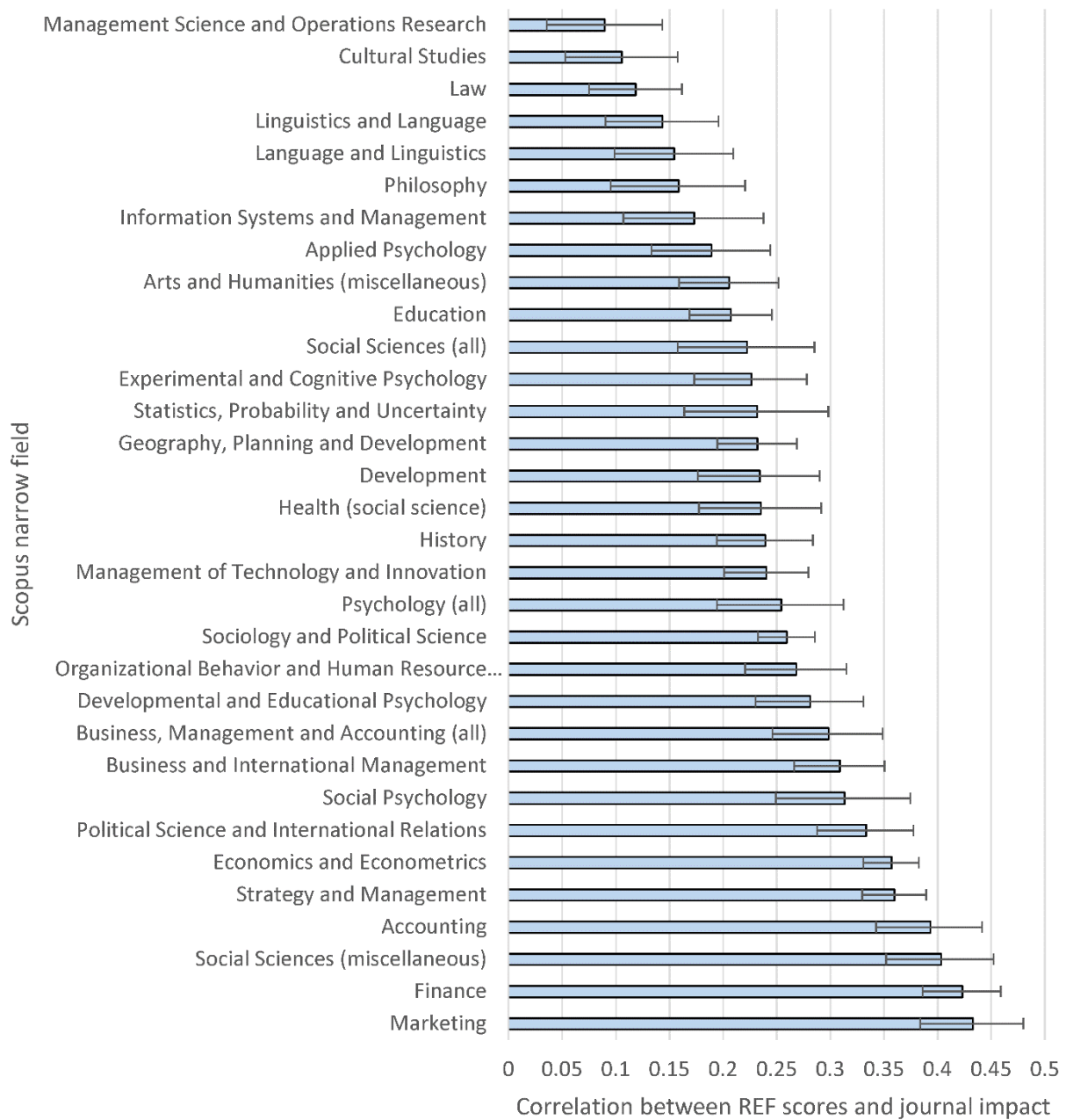

Figure 7. Spearman correlations by **narrow field** between average journal impact (MNLCS) for UK REF provisional score for UK REF2021 articles matched with a Scopus journal article published 2014-18 within a **Social Sciences** broad field. Error bars illustrate 95% confidence intervals. Qualification: At least 750 articles with REF scores.

# 4   Discussion

The results are limited by considering a single period (2014-18) and may change in the future as journals and fields evolve. They are also restricted to results from a single country, and other country evaluators may consider different criteria when judging the quality of an article (Taylor & Willett, 2017), such as its value for practical solutions. Whilst the UK REF is almost an ideal case in the sense of large-scale expert judgements by people explicitly told ignore the reputation of the publishing journal, individual sub-panel members in some UoAs may have disregarded this advice or have been subconsciously influenced, based on their own perceptions of their fields. The journal impact calculation is also a limitation. Since the MNLCS method used here was designed to be optimal for fair assessments of average journal impact in multi-disciplinary contexts, correlations may well be lower for the journal impact indicators available from Scopus and the Web of Science, especially because they do not use log normalisation, so can give misleading averages. A weakness of the MNLCS field normalisation component is that it relies on the Scopus narrow field categorisation scheme, which may be imperfect for this purpose.

The results mostly align with the REF2014 journal comparison from HEFCE (Wilsdon et al., 2015b), except that all correlations were positive for 2014-18, perhaps due to larger data set (almost five times more articles), and a much higher correlation was found for Economics and Econometrics (0.63 for 2008 rather than 0.5 for 2014-18). The difference may be due to the impact factor used if published impact factors are consulted and considered important to economists, to changes in the Scopus-indexed journal set, or to the evolution of economics as a discipline (e.g., changing methodological orientations: Cherrier & Svorenčík, 2018).

The results also broadly align with previous research. They align with Australian findings that expert rankings of journals positively correlate with journal impact for all 27 Scopus broad fields (Haddawy et al., 2016). They also tend to confirm numerous previous studies showing that the expert-judged value of a journal tends to correlate positively with its citation impact, including the stronger correlations for health-related fields and weaker correlations for the social sciences (Mahmood, 2017). They also confirm the relatively strong correlations between journal prestige and citation impact in psychology (Highhouse et al., 2020) and business Walters & Markgren, 2019). This study extends all these academic analyses by using the same period and method to allow comparisons across the whole of science.

In terms of practical applications, the weak to moderate correlations for business confirm that journal impact is not irrelevant in this field and that it is imperfect, supporting their minor role in a prominent business ranking (Kelly et al., 2013). Similarly, the results agree with the supporting role of impact factors in creating national lists of journals for evaluation purposes (Pölönen et al., 2021).

## 5 Conclusion

The results show that, at least for articles submitted to UK REF2021 and published 2014-18, higher quality articles tend to be published in higher impact journals in all REF UoAs (n=34), all broad fields (n=27), and nearly all narrow fields of science (n=94 shown), with the sole exception not being an academic field. The correlations are very weak (0.11) to moderate (0.43) for broad fields, and stronger (0.54) for Multidisciplinary, perhaps due to competitive generalist journals like *Science* and *Nature*. Weaker correlations may reflect non-hierarchical subjects, where journal specialty is more relevant than any journal prestige.

The lack of a strong correlation between article quality and average journal impact within specific fields (e.g., never above 0.5 for any UoA, never above 0.42 for any broad field, never above 0.54 for any large narrow field) confirms that journal impact is not ever an accurate proxy for the quality of individual articles. This result confirms DORA's advice to avoid using journal impact factors as a proxy for article quality, including for the fields in which the association is strongest. Nevertheless, since the correlation between article quality and journal impact is almost always positive, albeit frequently weak, it seems reasonable for scholars and evaluators to take journal citation rates into consideration when making decisions, especially when there is a lack of expertise, time or impartiality to fully evaluate individual articles or when only aggregate scores are needed for large sets of articles. Since there is not a simple rule about the fields in which journal impact is the least weak indicator of article quality, the graphs in this article may serve as a reference point to lookup the level of importance that may be attributed to journal impact in any given field.